\documentclass[traditabstract]{aa}
\usepackage{txfonts}
\usepackage{natbib}
\bibpunct{(}{)}{;}{a}{}{,}
\usepackage{graphicx}

\newcommand\epsoph{\ensuremath{\epsilon~\mathrm{Oph}}}
\newcommand\muhz{\ensuremath{\mu\mathrm{Hz}}}
\newcommand\ea{et al.}
\newcommand{\hrd}{HR diagram}
\newcommand{\msun}{\ensuremath{M_\odot}}
\newcommand{\rsun}{\ensuremath{R_\odot}}
\newcommand{\lsun}{\ensuremath{L_\odot}}
\newcommand{\febyh}{\ensuremath{\mathrm{[Fe/H]}}}
\newcommand{\teff}{\ensuremath{T_\mathrm{eff}}}
\newcommand{\lteff}{\ensuremath{\log T_\mathrm{eff}}}
\newcommand{\chisqnu}{\ensuremath{\chi_\nu^2}}
\newcommand{\chisqdeltanu}{\ensuremath{\chi_{\Delta\nu}^2}}

\def\myfigure#1#2#3#4{
	\begin{figure#4}
	\resizebox{\hsize}{!}{\includegraphics{#1}}
	\caption{#2 \label{#3}}
	\end{figure#4}
}

\begin{document}

\title{Asteroseismology and interferometry of the red giant star \epsoph}

\author{ 
     A.~Mazumdar\inst{1,2,3}
\and A.~M\'erand\inst{4,5} 
\and P.~Demarque\inst{2}
\and P.~Kervella\inst{6}
\and C.~Barban\inst{6,1}
\and F.~Baudin\inst{7} 
\and V.~Coud\'e~du~Foresto\inst{6} 
\and C.~Farrington\inst{5}
\and P.~J.~Goldfinger\inst{5} 
\and M.-J.~Goupil\inst{6}
\and E.~Josselin\inst{8}
\and R.~Kuschnig\inst{9}
\and H.~A.~McAlister\inst{5} 
\and J.~Matthews\inst{10}
\and S.~T.~Ridgway\inst{11} 
\and J.~Sturmann\inst{5} 
\and L.~Sturmann\inst{5} 
\and T.~A.~ten~Brummelaar\inst{5}
\and N.~Turner\inst{5} 
}

\offprints{anwesh@tifr.res.in}

\institute{
Instituut voor Sterrenkunde, Katholieke Universiteit,
200D Celestijnenlaan, 3001 Leuven, Belgium
\and 
Astronomy Department, Yale University,
PO Box 208101, New Haven CT 06520-8101, USA
\and
Homi Bhabha Centre for Science Education, TIFR,
V.~N.~Purav Marg, Mankhurd, Mumbai 400088, India
\and
European Southern Observatory, Alonso de C\'ordova 3107, Casilla
19001, Santiago 19, Chile 
\and
Center for High Angular Resolution Astronomy, Georgia State
University, PO Box 3965, Atlanta, Georgia 30302-3965, USA 
\and
LESIA, Observatoire de Paris, CNRS\,UMR\,8109, UPMC, Universit\'e
Paris Diderot, 5 Place Jules Janssen, 92195 Meudon, France 
\and
Institut d'Astrophysique Spatiale, CNRS/Universit\'{e} 
Paris XI UMR 8617, 91405 Orsay Cedex, France
\and
GRAAL, Universit\'e Montpellier II, CNRS\,UMR\,5024, 34095 Montpellier 
Cedex 05, France
\and
Institut f\"{u}r Astronomie, Universit\"{a}t Wien, 
T\"{u}rkenschanzstrasse 17, 1180 Vienna, Austria
\and
Department of Physics \& Astronomy, University of British Columbia, 
6224 Agricultural Road, Vancouver V6T 1Z1, Canada
\and
National Optical Astronomy Observatories, 950 North Cherry Avenue,
Tucson, AZ 85719, USA 
}

\date{}

\abstract{
The GIII red giant star \epsoph\ has been found to exhibit several modes
of oscillation by the MOST mission.  We interpret the observed
frequencies of oscillation in terms of theoretical radial $p$-mode
frequencies of stellar models.  Evolutionary models of this star, in
both shell H-burning and core He-burning phases of evolution, are
constructed using as constraints a combination of measurements from
classical ground-based observations (for luminosity, temperature, and
chemical composition) and seismic observations from MOST.  Radial
frequencies of models in either evolutionary phase can reproduce the
observed frequency spectrum of \epsoph\ almost equally well. The
best-fit models indicate a mass in the range of $1.85 \pm 0.05 \msun$
with radius of $10.55 \pm 0.15 \rsun$.  We also obtain an independent
estimate of the radius of \epsoph\ using high accuracy interferometric
observations in the infrared $K'$ band, using the CHARA/FLUOR
instrument.  The measured limb darkened disk angular diameter of
\epsoph\ is $2.961 \pm 0.007$\,mas. Together with the \emph{Hipparcos}
parallax, this translates into a photospheric radius of $R = 10.39 \pm
0.07\rsun$.  The radius obtained from the asteroseismic analysis matches
the interferometric value quite closely even though the radius was not
constrained during the modelling.
}

\keywords{Stars: individual: $\epsilon$~Ophiuchi -- Stars: oscillations 
-- Stars: interiors -- Stars: fundamental parameters -- 
Techniques: interferometric}

\maketitle

\section{Introduction}
\label{sec:intro}

Asteroseismology of red giant stars has, in recent years, taken a leap
forward with the discovery of pulsations in several G- and K-type giant
stars, both from the ground \citep{frandsen02,deridder06} and from space
\citep{barban07,kal08b,hekker08}. Oscillations in the G giant
\object{\epsoph}\ (\object{HD 146791}, \object{HR 6075}, \object{HIP
79882}) was first detected in spectroscopic observations from the ground
\citep{deridder06}, although the average large separation could not be
distinguished between two possible values due to the daily alias
problem. Subsequent observations by the MOST satellite \citep{walker03}
led to the discovery of at least 9 radial modes with an average large
separation of $5.3\pm 0.1\,\muhz$ \citep{barban07}.

This work makes an attempt to interpret the observed frequencies of
\epsoph\ in terms of adiabatic oscillation modes of stellar models in
the relevant part of the \hrd. We construct red giant models in which
the luminosity is provided by either hydrogen burning in a shell outside
the helium core, or both shell hydrogen burning and core helium burning.
We make quantitative comparisons of these models to the MOST frequencies
of \epsoph\ to determine the stellar parameters like mass, age, radius,
and chemical composition.  \citet{kal08a} have earlier presented stellar
models in the shell hydrogen-burning phase for \epsoph, based on
asteroseismic data.  A similar study of oscillations in the red giant
\object{$\xi$~Hya} in terms of helium burning models was carried out by
\citet{jcd04}.

Interferometric measurements of stellar radii are particularly
discriminating for models, in particular when combined with
asteroseismic frequencies, as noticed, for instance, by
\citet{creevey07} and \citet{cunha07}. In this work we report a new
interferometric determination of the radius of \epsoph. While the direct
measurement of the radius of a red giant is a useful result in itself,
for \epsoph\ it provides the first opportunity of testing the relevance
of theoretical models for red giants which have been calibrated with
asteroseismic input. In this study, neither did we use the
interferometric radius as an input to the modelling, nor did the
interferometric analysis draw upon the asteroseismic information in any
way. Thus the radius of the stellar models that fit the seismic data
best can be tested against the independently measured interferometric
radius. 

In Sect.~\ref{sec:models} we describe the details of the stellar models
that we have constructed and in Sect.~\ref{sec:freq} we compare the
theoretical frequencies obtained from these models with the observed
MOST frequencies.  In Sect.~\ref{sec:interf}, we present our new
interferometric measurement of the angular diameter of \epsoph. In
Sect.~\ref{sec:discuss} we compare the radii of our best seismic models
with the interferometric measurement, and discuss our results with
similar studies carried out earlier.

\section{Stellar models and theoretical frequencies}
\label{sec:models}

We constructed a grid of stellar models with various input parameters
using the Yale Rotating Evolutionary Code (YREC) \citep{gue92}. This
code is capable of producing consistent stellar models for low mass
giant stars both in the shell H-burning phase and the core He-burning
phase \citep{yrec}. We describe these two sets of models below. The
radial and nonradial pulsation frequencies of each model are calculated
by the oscillation code JIG \citep{jig}. 

\subsection{Input physics}
\label{sec:models-physics}

The models use the latest OPAL equation of state \citep{opaleos05} and
OPAL opacities \citep{opalop95}, supplemented by the low temperature
opacities of \citet{ferg05}. The nuclear reaction rates from
\citet{bahcall92} are used. Diffusion of helium and heavy elements have
been ignored in the post-main sequence phase of evolution. This is
reasonable, since dredge-up by the deep convective envelope present in
red giants would mask any effect of diffusion of elements in early
phases. The current treatment of convection in the models is through the
standard mixing length theory \citep{bohm58}, which does not properly
include the effects of turbulence in the outer layers, and this might
substantially affect the frequencies of oscillation \citep[see
e.g.,][]{straka07}. Fortunately, the uncertainty induced in the large
separations is much less than in the actual frequencies themselves.
Mass loss on the giant branch was not included in the calculations. Most
of the mass loss is believed to take place quiescently as the star
approaches the tip of the giant branch, as is the case in the commonly
adopted \citet{reimers77} formulation \citep{yi03}.  Such mass loss does
not affect the thermodynamics of the deep interior appreciably.  Most
importantly, it takes place at high luminosities, beyond the luminosity
of \epsoph\ on the giant branch.  The neutrino losses in the core were
taken from the work of \citet{itoh89}.

\subsection{Range of parameters}
\label{sec:models-param}

The range of input parameters chosen for the models is dictated by the
position of \epsoph\ on the \hrd\ and its estimated chemical
composition.  The values of effective temperature ($\lteff =
[3.680,3.698]$), luminosity ($\log L/\lsun = [1.732,1.806]$), and
metallicity ($\febyh = [-0.07,-0.17]$) are adopted from
\citet{deridder06}, who have already carried out a detailed survey of
these parameters from the literature.  Tracks were constructed for
different ($Y_0$, $Z_0$) combinations with $Y_0$ ranging from 0.255 to
0.280, and $Z_0$ ranging from 0.012 to 0.015, corresponding to $\febyh =
(-0.07, -0.17)$.  We adopted the solar abundances as given by
\citet{gs98} in converting \febyh\ to $Z$ values.

For each of these combinations, two values of the mixing length
parameter $\alpha$ (the ratio of mixing length to local pressure scale
height) have been used: 1.6 and 1.8.  Note that using a similar version
of YREC and input physics,  \citet{kal08a} used $\alpha = 1.74$ in their
study of \epsoph.  This is the value of $\alpha$ adopted by \citet{yi03}
for the standard solar model calibration.  The radii of red giant models
depend sensitively on the choice of $\alpha$. 

For each ($Y_0$,$Z_0$,$\alpha$), the mass has been varied between 1.8
and 2.4\msun\ to check for overlap in the box.  In most models,
overshoot at the edge of the convective core present on the main
sequence was assumed to have negligible effect on the advanced
evolution.  A few models were constructed with overshoot of  0.2 times
the pressure scale height at the convective core edge.  Because the size
of the convective core is not too large in this mass range, the overall
structural effect of core overshoot  is modest; but  evolutionary
timescales are slightly increased when core overshoot is taken into
account \citep{dem04}. 

\subsection{Shell H-burning models}
\label{sec:models-shellh}

Our first set of models for \epsoph\ are on the ascending red giant
branch. These models are characterised by an inert helium core
surrounded by a thin hydrogen burning shell. The mass of the shell
varies between $0.012\msun$ and $0.0007\msun$ depending on the mass and
age. The mass of the shell decreases as the star ascends the red giant
branch.  We concentrate on models that lie within the \epsoph\ box on
the \hrd. For each evolutionary track that traverses the box, several
models at slightly different ages are constructed so as to span the box.
The theoretical frequencies of these models are compared to the observed
frequencies of \epsoph\ in Sect.~\ref{sec:freq}.

\myfigure{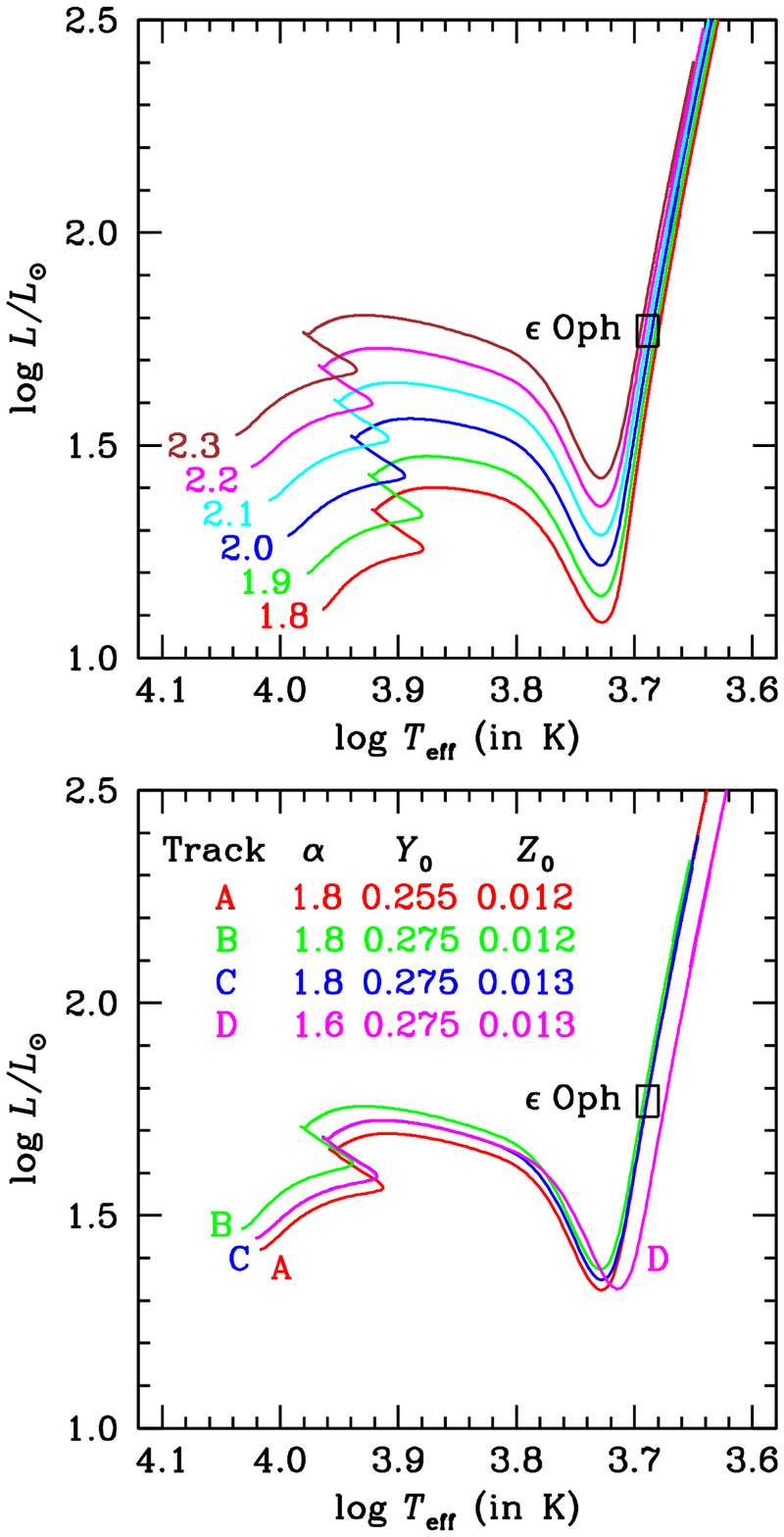}
{Stellar evolutionary tracks from the ZAMS to the shell H-burning RGB
are shown with respect to the position of \epsoph\ on the \hrd.  The
{\it top} panel shows tracks for different masses ($M/\msun$ indicated)
with identical initial chemical composition ($Y_0=0.270$, $Z_0=0.012$)
and mixing length ($\alpha=1.8$).  The {\it bottom} panel shows tracks
with the same mass ($M=2.2\msun$) but with different values of $Y_0$,
$Z_0$, and $\alpha$.} 
{fig:hrd-rgb}
{} 

Each evolutionary track was started in the pre-main sequence phase and
evolved continuously through core hydrogen burning and eventually shell
hydrogen burning along the giant branch. Some of the tracks are plotted
in the \hrd\ in Fig.~\ref{fig:hrd-rgb}, together with the \epsoph\ error
box. The size of this box is such that the range of mass of models with
a given set of ($Y_0$, $Z_0$) and $\alpha$ values that pass through the
box is $\sim 0.5\msun$ (see top panel of Fig.~\ref{fig:hrd-rgb}).
Typically, the mass lies between $1.8\msun$ and $2.4\msun$, depending on
the values of the other parameters. The shift in the tracks with these
parameters is also significant (see bottom panel of
Fig.~\ref{fig:hrd-rgb}).  Since the outer convective layer of stars in
this mass range is extremely thin during the main sequence phase, the
tracks with different values of $\alpha$ are almost identical in that
phase (e.g., tracks C and D in Fig.~\ref{fig:hrd-rgb}). But the shift of
the track on the giant branch is significant because of the extended
convective envelope. In most cases in this study, however, frequencies
of models inside the \epsoph\ box with $\alpha = 1.6$ had a poor match
with the observed frequencies. Since the tracks move redwards with
decreasing $\alpha$, this can be traced to larger mass (and hence larger
radii inside the box) of the $\alpha=1.6$ models compared to the
$\alpha=1.8$ models. Keeping in mind the adopted range of \febyh\ for
the star, the metallicity of the initial model can only be varied
between 0.012 and 0.015 with corresponding appropriate change of initial
helium abundance between 0.255 and 0.280 to span the entire \epsoph\ box
on the \hrd.

\subsection{Core He-burning models}
\label{sec:models-corehe}

The second set of models are in a later stage of evolution than the
first. These models have helium burning in the core of the star.
Hydrogen burning in a thin shell outside the core is also present.
Considering the adopted metallicity value for \epsoph, these models
represent the so-called ``red clump'' stars, rather than metal-poor
horizontal branch stars. Indeed, the evolutionary tracks of the models
that we constructed lie very close to the red giant branch, and
therefore, overlap with the error box of \epsoph\ on the \hrd.
Fig.~\ref{fig:hrd-hb} illustrates these tracks in the \hrd.
\myfigure{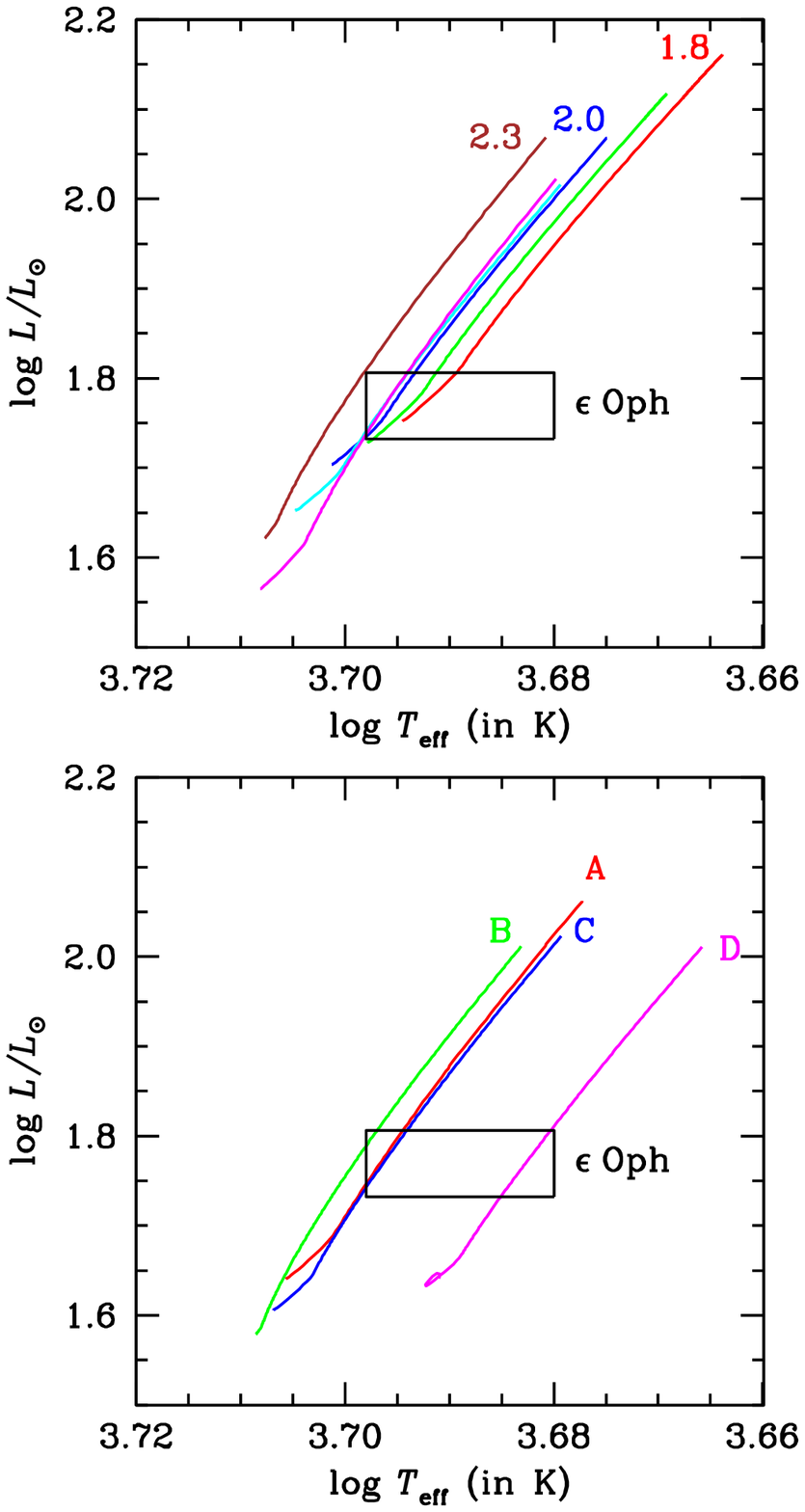}
{Stellar evolutionary tracks from the ZAHB to the  end of He-burning
main sequence are shown with respect to the position of \epsoph\ on the
\hrd. The direction of evolution is from the lower left to the upper
right corner of the graphs. The tracks in each panel are the successors
of those in Fig.~\ref{fig:hrd-rgb}. The {\it top} panel shows tracks for
different masses ($M/\msun$ indicated) with identical values of initial
chemical composition and mixing length.  The {\it bottom} panel shows
tracks with the same mass ($M=2.2\msun$) but with different values of
$Y_0$, $Z_0$, and $\alpha$ (given in Fig.~\ref{fig:hrd-rgb}).} 
{fig:hrd-hb}
{} 

For the range of chemical composition used in our models, it turns out
that the helium ignition in the core at the tip of the red giant branch
takes place under degenerate conditions for the lower mass stars
($M/\msun \la 2.1$). This is the well-known ``helium flash'' mechanism
first studied by \citet{sch62}.

For stars of slightly higher mass ($M/\msun \ga 2.1$), however, the core
is not degenerate at the instant helium burning temperatures are
reached, and therefore, helium ignition takes place in a controlled
fashion. In this latter case, it is numerically easy to continue the
evolution of the model past the tip of the red giant branch, and onto
the red clump phase.

For the helium flash scenario, however, the numerical stability of the
evolution code at the tip of the giant branch is far less, and only a
computationally expensive algorithm involving subtle handling of various
parameters can guide the model past the runaway helium ignition process
and settle it onto a stable phase of helium burning \citep{dem71}.
\citet{ptc04} have demonstrated that even for stars which undergo a
violent helium flash, the subsequent evolution of a model on the
helium-burning main sequence, i.e., once stable core helium burning has
been established, is not very sensitive to the prior history of helium
ignition.  Specifically, the behaviour of models which have been evolved
from appropriate zero age horizontal branch (ZAHB) models with quiescent
helium burning in the core is remarkably similar to that of models which
have been actually evolved through the helium flash phenomenon. Of
course, the make-up of the ZAHB model is crucial -- it must reflect the
properties of a model that has settled on the helium-burning main
sequence after having gone through the helium flash. The critical factor
in this starting model, apart from the total mass and the chemical
composition at the tip of the giant branch, is the mass of the helium
core.  Hydrodynamical studies in 2D \citep{col83} and 3D \citep{moc08}
confirm this picture, except for possible  second order mixing effects
due to turbulent overshoot at the convective-radiative interface, which
cannot, at this point, be estimated precisely.  

For the low mass models we have, therefore, followed the approach of
circumventing the numerical difficulties encountered in handling the
helium flash, as done by most authors \citep[e.g.,][]{lee90,swe87}.  For
each mass and chemical composition evolution was continued on the red
giant branch till the onset of helium flash.  The evolution of the red
clump model was then re-started from a ZAHB model of the same total mass
with identical chemical composition and helium core mass as the
corresponding model at the onset of helium flash at the tip of the red
giant branch. 

Fig.~\ref{fig:heflash} illustrates the behaviour of the central regions
of a 2\msun\ model near the tip of the giant branch, where helium
ignition takes place.  The four panels  describe the change, as a
function of time, or equivalently, as a function of maximum temperature
in the star, $T_\mathrm{max}$, of the following quantities: the mass
contained interior to the shell at $T_\mathrm{max}$,
$m(T_\mathrm{max})/\msun$ , the degeneracy parameter $\eta$, the energy
generation rate due to the triple-alpha reaction, $\epsilon(3\alpha)$,
and finally the mass of the helium core, $m_\mathrm{He-core}/\msun$.
Note that the shell with the highest temperature is  not central (i.e.,
$m(T_\mathrm{max})/\msun \neq 0$) and changes with time.  This is
because of neutrino losses, which are most effective  at the higher
densities near the centre. As the degeneracy increases, neutrino cooling
causes such an inversion of the temperature profile near the centre till
the helium ignition temperature is reached. Such off-centre ignition of
helium is typical in degenerate helium cores. The energy released in the
helium ignition reduces the degeneracy, and the shell of maximum
temperature moves back to the centre of the star.  The degeneracy
parameter $\eta$ is a measure of the degree of degeneracy of the
electron gas.  It is a dimensionless parameter used to quantify the
relationship between the electron density and pressure in a partially
degenerate electron gas.  The detailed formalism used in the models is
that  described by \citet[][p.~64]{clayton68}. We note, while inspecting
Fig.~3, that the quantity $\eta$ varies between $-\infty$ in the ideal
gas case and $+\infty$ in a fully degenerate Fermi-Dirac gas.  The mass
of the helium core keeps increasing throughout the red giant phase due
to hydrogen burning in the shell immediately above it till the onset of
helium burning. The maximum mass of a degenerate helium core at helium
flash is typically $\sim 0.45\msun$, irrespective of the total mass of
the star, and depends slightly on the other stellar parameters. 

\myfigure{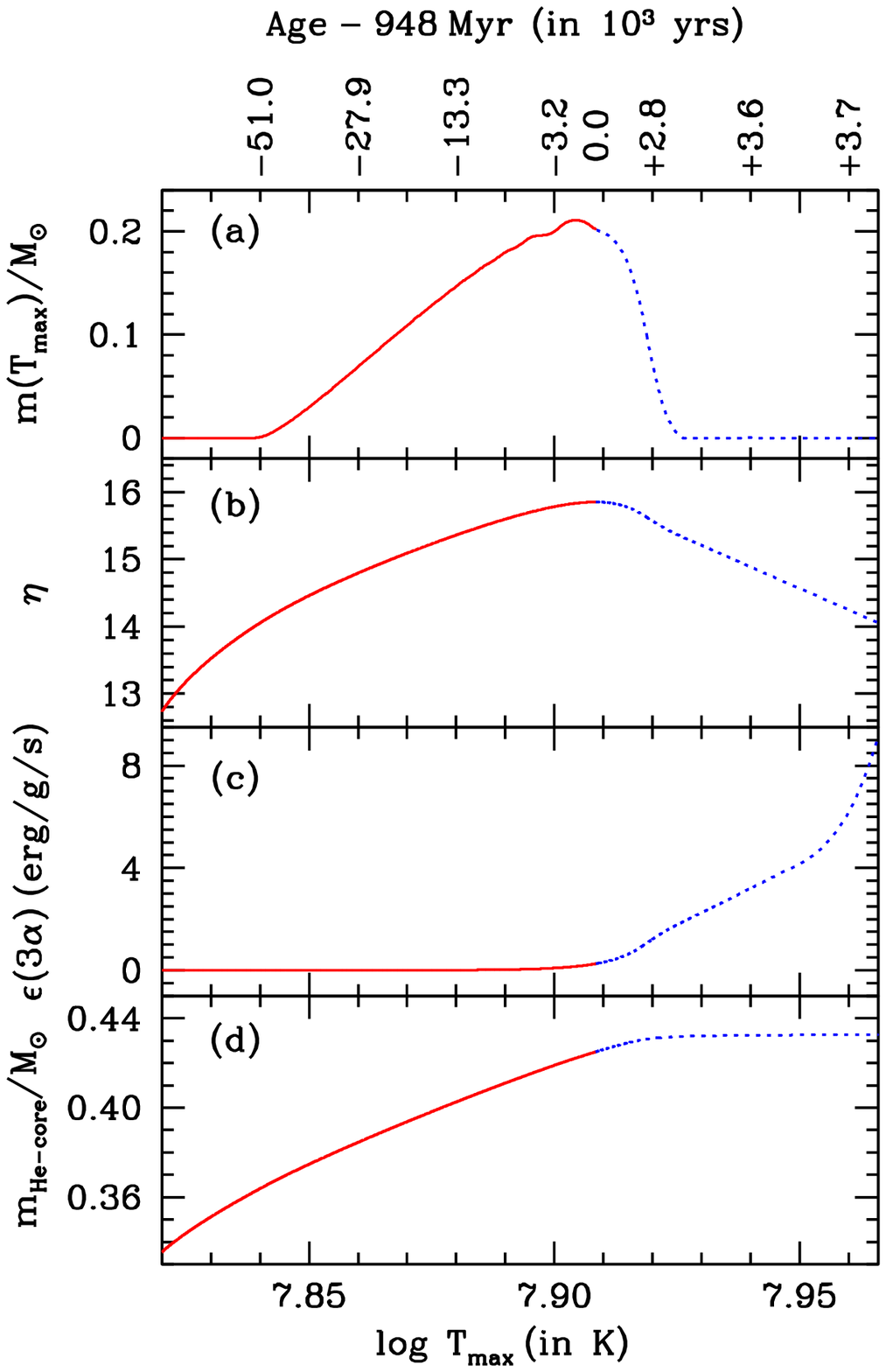}
{The change, as a function of time (upper $x$-axis), or equivalently as
a function of maximum temperature in the star, $T_\mathrm{max}$ (lower
$x$-axis), of (a) the mass interior to the shell at $T_\mathrm{max}$,
$m(T_\mathrm{max})/\msun$, (b) the degeneracy parameter, $\eta$, (c) the
energy generation due to the triple-alpha reaction, $\epsilon(3\alpha)$,
and (d) the mass of the helium core, $m_\mathrm{He-core}/\msun$, close
to the onset of He-burning of a 2\msun\ star are shown. The {\it red
solid line} denotes the pre-He-ignition phase, while the {\it blue
dotted line} shows the post-He-ignition phase.} 
{fig:heflash} 
{}

The actual age of the model on the helium-burning main sequence cannot,
of course, be assigned accurately due to the ``missing'' period of the
helium flash. However, the total duration of the helium flash phenomenon
and the subsequent stabilisation of the star on the helium-burning main
sequence is only $\sim 1.5$\,Myr \citep{ptc04}, and hence the
uncertainty in the age in our helium-burning models is quite small.  

For slightly higher mass models ($M/\msun \ga 2.2$), the evolution of
the star is followed continuously from the pre-main sequence stage till
the red clump stage. Since the helium ignition at the tip of the giant
branch occurs under non-degenerate conditions, there are no numerical
problems in such cases.  The transition mass from violent to quiescent
helium ignition is a function of chemical composition.  It has been
studied in detail by \citet{swe89}. More recent illustration is found in
the tracks of \citet{yi03}, which were constructed with a similar
version of YREC \citep{yrec}.

There is an uncertainty in the age of the core helium-burning models due
to the assumption of no mass loss on the red giant branch. Since the
amount of mass lost in the giant phase is not known, the mass of the
ZAHB model cannot be assigned accurately. Therefore, the ages of
helium-burning models given in Tables~\ref{tab:bestfit}
and~\ref{tab:best-rad} are, in fact, upper limits to the real ages.

\subsection{Timescales of evolution}
\label{sec:models-timesc}

\myfigure{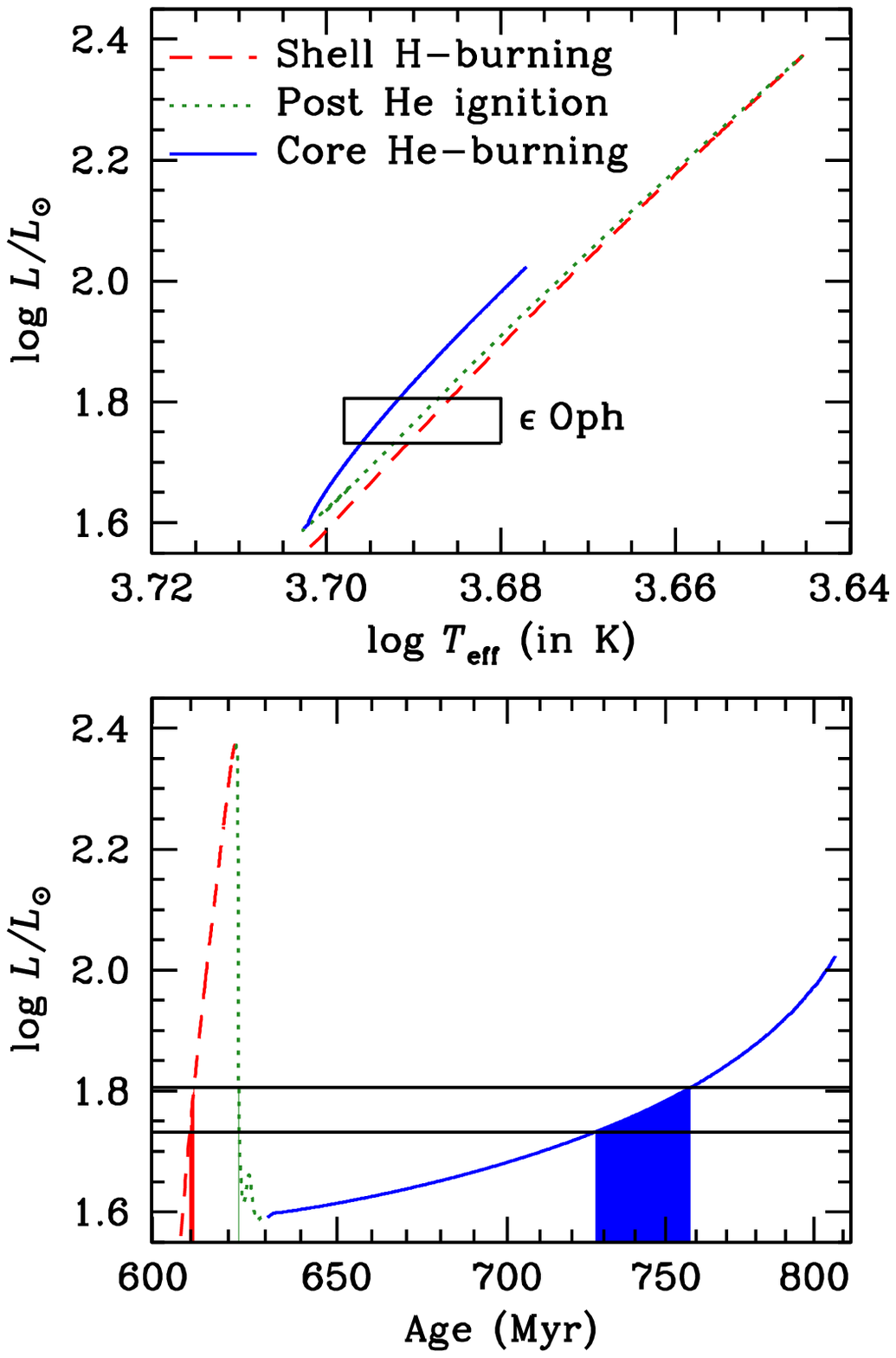}
{The different timescales of the stellar evolutionary tracks in each
crossing of the \epsoph\ box for the shell H-burning, post He-ignition,
and core He-burning main sequence are shown. In the {\it top} panel the
full evolutionary track of a 2.3\msun\ model is shown on the \hrd. The
{\it bottom} panel shows the luminosity of the star as a function of its
age.  The luminosity limits of \epsoph\ are indicated by the two
horizontal lines, and the time spent by the star within these limits in
each phase are indicated by the projections on the {\it x}-axis.}
{fig:hrd-timesc}
{}

A star of a given mass might cross the \epsoph\ errorbox on the \hrd\
three times -- once upwards during shell H-burning, once downwards
during the stabilisation of the star just after He ignition, and once
again upwards during the phase of stable core He-burning. The timescales
of evolution in these three phases are quite different, and consequently
the time spent inside the \epsoph\ box differs vastly.
Fig.~\ref{fig:hrd-timesc} illustrates the timescales of evolution in
each crossing of the box for a 2.3\msun\ star. During the shell
H-burning phase, it spends nearly 1.09\,Myrs inside the box. After He
ignition in the core, it spends only 0.17\,Myr during the rapid settling
towards the He-burning main sequence, and finally it spends 30.59\,Myrs
during the stable core He-burning phase. Similar timescales are found
for other stars in this mass range. The time spent during the shell
H-burning phase is typically $\sim 20$ times shorter than that during
the core He-burning phase. Thus the probabilities of \epsoph\ being in
the corresponding phases of evolution are in the same ratio.

\section{Comparison of theoretical and observed frequencies}
\label{sec:freq}

The theoretical frequencies of the stellar models were compared with the
MOST data on \epsoph. Typically, in asteroseismic modelling studies, the
comparison between a stellar model and the observed frequency data is
carried out in terms of frequency separations, especially the large
frequency separations, rather than the frequencies themselves.  This is
done to eliminate the uncertainty in the theoretical absolute
frequencies due to inadequate modelling of the stellar surface layers.
However, this is possible only in the happy circumstance of detection of
a series of frequencies of the same degree and successive radial orders,
for which the large separations can be determined. For \epsoph,
\citet{barban07} indeed provide the frequencies of 9 successive radial
order modes.  Thus it is possible to match the observed large
separations with the theoretical values from the models. However, as an
additional comparison, we also match the absolute frequencies of radial
modes of our stellar models with the MOST data. 

\begin{table*}
\caption{
Stellar parameters for the models with lowest \chisqdeltanu\ and
\chisqnu\ values in either shell H-burning or core He-burning phase are
given.  
\label{tab:bestfit}
}
\begin{tabular}{clcccccccccc}
\hline
\hline
Minimised by & Evolutionary phase & $Y_0$ & $Z_0$ & $\alpha$ & $M/\msun$ & Age (Myr) 
& \lteff & $\log L/\lsun$ & $R/\rsun$ 
& \chisqnu  & \chisqdeltanu  \\
\hline
\chisqdeltanu
 & Shell H-burning & 0.260 & 0.012 & 1.8 & 1.9 & 1092 & 3.682 & 1.732 &
10.54  & 10.68\phantom{1} & 0.62\\
 & Core He-burning & 0.265 & 0.012 & 1.8 & 1.8 & 1326 & 3.692 & 1.766 &
10.51  & 0.55 & 0.59\\
\hline
\chisqnu 
 & Shell H-burning & 0.260 & 0.013 & 1.8 & 1.9 & 1063 & 3.681 & 1.741 &
10.71  & 0.47 & 0.63\\
 & Core He-burning & 0.255 & 0.012 & 1.8 & 1.9 & 1173 & 3.692 & 1.780 &
10.69  & 0.51 & 0.68\\
\hline
\end{tabular}
\end{table*}

For each comparison, a reduced $\chi^2$ value is computed as
\begin{equation}
\chisqnu = \frac{1}{N}\sum_{i=1}^N \left[\frac{\nu_\mathrm{MOST} -
\nu_\mathrm{model}}{\delta\nu_\mathrm{MOST}}\right]^2
\end{equation}  
\begin{equation}
\chisqdeltanu = \frac{1}{N-1}\sum_{i=2}^{N}
\left[\frac{\Delta\nu_\mathrm{MOST} -
\Delta\nu_\mathrm{model}}{\delta\Delta\nu_\mathrm{MOST}}\right]^2
\end{equation}  
where $\nu$ and $\Delta\nu$ represent the frequency and large separation
respectively, and $N$ is the number of observed modes compared (maximum
of 9).  $\delta\nu_\mathrm{MOST}$ is the measured error in the MOST
frequency and $\delta\Delta\nu_\mathrm{MOST}$ is the error in the large
separation computed by adding the adjacent frequency errors in
quadrature. Thus for each model, values of \chisqnu\ and \chisqdeltanu\
can be computed. A small value of \chisqnu\ would usually imply a small
value for \chisqdeltanu, but the reverse is not necessarily true. It is
possible to find models for which the large separations are quite close
to the observed values, but all the frequencies are shifted by a nearly
constant amount. Given the inadequacies of input physics in the models,
especially in the rarefied outer convective layers, the absolute values
of the theoretical frequencies cannot be trusted too much. The large
separations, on the other hand, would be relatively free from such
ambiguities, and would reflect the overall structure of the star better.
This is why we put a greater importance on minimising \chisqdeltanu,
rather than \chisqnu, in choosing our best model for \epsoph. It turns
out, however, that the models with lowest \chisqdeltanu\ have reasonably
small values for \chisqnu\ as well in most cases.  \citet{kal08a} have
used absolute frequencies in their comparison of models and observations
of \epsoph.

The best-fitting models in both shell H-burning and core He-burning
phases can reproduce the observed large separations fairly well. This is
illustrated in Fig~\ref{fig:lsep}.  However, it turns out that the large
separation corresponding to one of the observed modes at
$\nu_\mathrm{MOST}=62.871\,\muhz$ is nearly $2\sigma$ away from the
theoretical value in all our models. This data point is always the
largest contributor to  \chisqnu\ and \chisqdeltanu. We have ignored
this point while choosing our best model.
\myfigure{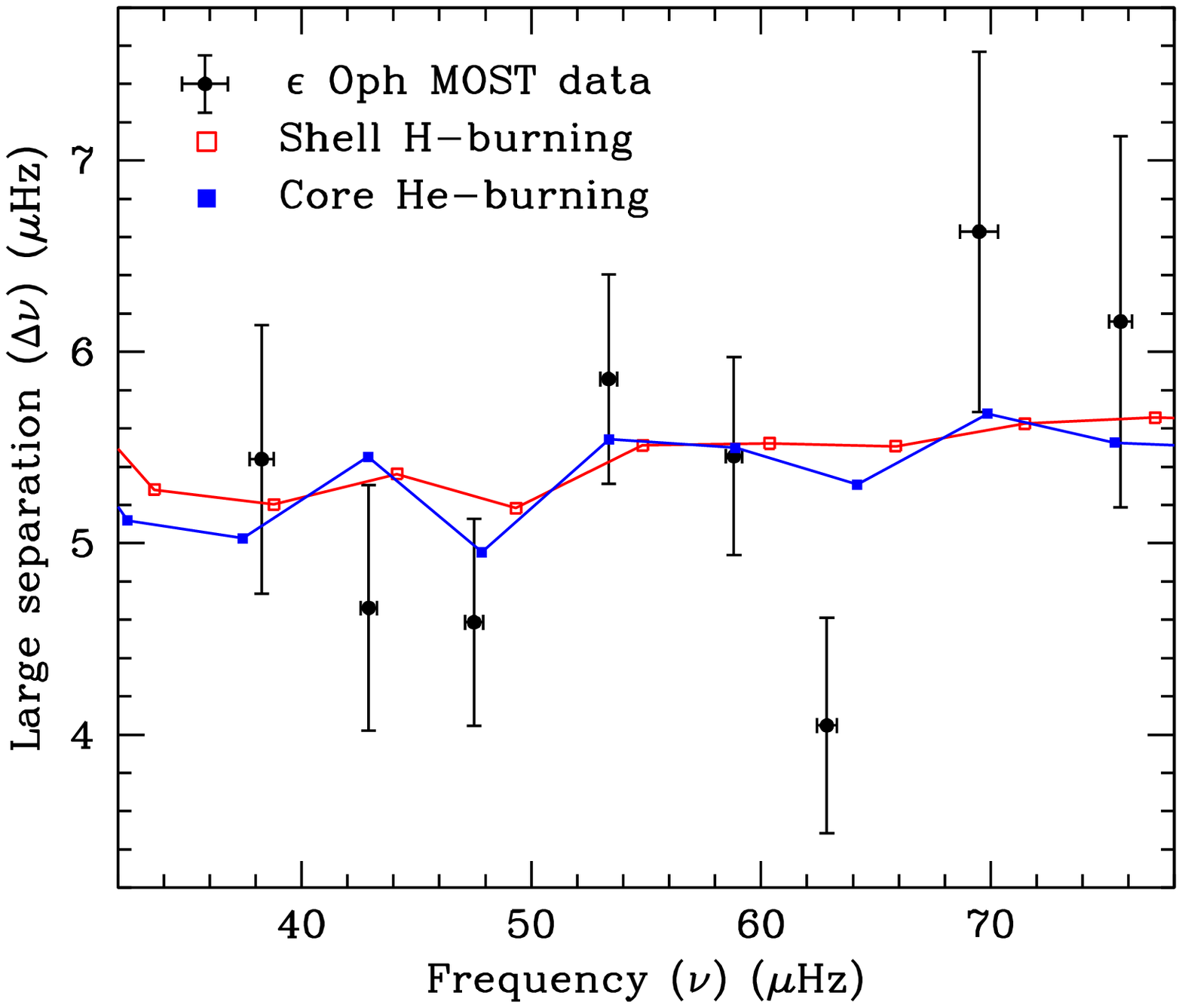}
{The large frequency separations of best-fitting radial modes of stellar
models in the shell H-burning ({\it red open squares}) and core
He-burning ({\it blue filled squares}) phases inside the \epsoph\
errorbox are shown in comparison with the observed separations of
\epsoph. The parameters of these models are given in
Table~\ref{tab:bestfit}. These models are chosen on the basis of lowest
\chisqdeltanu\ criterion.}
{fig:lsep}
{}

The parameters of the models with the lowest \chisqdeltanu\ and
\chisqnu\ values are listed in Table~\ref{tab:bestfit}. Notice that the
model with least \chisqdeltanu\ in the shell H-burning phase has a
significantly large value of \chisqnu\ due to an overall shift in the
absolute frequencies. Given one criterion for comparison (either
\chisqdeltanu\ or \chisqnu), it is clear that the models in the shell
H-burning phase fit the data almost equally well as those in the core
He-burning phase. Thus, the present data is unable to distinguish
between these two phases of stellar evolution.  However, as discussed in
Sect.~\ref{sec:models-timesc}, the likelihood of the star being in the
core He-burning phase is greater than it being in the shell H-burning
phase.

Based on the best match between observed and model large separations,
the models indicate very similar parameters for both phases of
evolution. We estimate the stellar parameters from not only the models
with lowest $\chi^2$, but actually all models that have $\chi^2$ values
within $50\%$ of the least $\chi^2$.  The mass is estimated to be $1.85
\pm 0.05 \msun$, while the metallicity of the best models are in the
range of $0.0125 \pm 0.0005$.  The radius lies in the range of $10.55
\pm 0.15 \rsun$. The radius of \epsoph, however, can be measured
independently through interferometry, as described in the next section.

\section{Interferometric measurements}
\label{sec:interf}

\subsection{Instrumental setup}

We observed \epsoph\ in July 2006 at the CHARA Array
\citep{tenbrummelaar05} using FLUOR, the Fiber Linked Unit for Optical
Recombination \citep{coude03}. This instrument is equipped with a near
infrared $K'$ band filter ($1.9 \leq \lambda \leq 2.35\,\mu\mathrm{m}$).
We extracted the instrumental visibilities from the raw data using the
FLUOR data reduction software \citep{coude97,kervella04,merand06}.  For
all the reported observations, we used the CHARA baselines S2-W2, with
ground lengths of 177\,m, which is mostly a north-south baseline in
orientation. The calibrator stars were chosen in the catalogue compiled
by \citet{merand05}, using criteria defined by these authors
(Table~\ref{tab:calibrators}). They were observed immediately before or
after our targets in order to monitor the interferometric transfer
function of the instrument. For a more detailed description of the
observing procedure and the error propagation, the interested reader is
referred to \citet{kervella08} and \citet{perrin03}, respectively. The
resulting calibrated squared visibilities are listed in
Table~\ref{tab:visib-table}.

\begin{table}
\caption{
Calibrators used for \epsoph. The limb darkened angular diameter
$\theta_{\rm LD}$ are given in milliarcseconds (mas). The angular
separation $\gamma$ between each calibrator and \epsoph\ is given in the
last column, in degrees.
\label{tab:calibrators}
}
\begin{tabular}{lccccc}
\hline \hline
Star & $m_V$ & $m_{Ks}$ & Spect.  & $\theta_{\rm LD}$ (mas) & $\gamma$\,($^\circ$) \\
\hline
\noalign{\smallskip}
\object{HR 145085} & 5.9 & 2.4 & K5III & $1.677\pm0.022$ & 8 \\
\object{HD 162468} & 6.2 & 3.2 & K1III & $1.154\pm0.015$ & 28 \\ 
\object{HD 166460} & 5.5 & 2.6 & K2III & $1.439\pm0.018$ & 29 \\
\hline
\end{tabular}
\end{table}

\subsection{Angular diameter measurement and precision}

In order to accurately measure \epsoph\ angular diameters, we used many
known stellar calibrators and we repeated the observation on two
separate and consecutive nights. Since, in the end, the number of
visibility will always be small, statistically speaking, the final
confidence on the precision will rely more on the repeatability of the
the result and the consistency between the stellar calibrators.

Our observation strategy was designed to give maximum precision and
confidence to our results. As a result, we achieve 
\begin{itemize}
\item the repeatability of the result. The first night gives
$\theta_\mathrm{UD}=2.881\pm0.006$\,mas, with a reduced $\chi^2$ of 0.9;
the second night $\theta_\mathrm{UD}=2.891\pm0.005$\,mas, with a reduced
$\chi^2$ of 0.3. This is consistent at the $0.01$\,mas level.
\item the use of multiple slightly resolved calibrators of various
sizes. Indeed, if we assume we have an overall bias in the angular
diameter estimation of the calibrators, it is going to lead to a
differential calibrated visibility bias, depending on the size of the
calibrator. For example, if we multiply the diameters of our calibrators
by 1.05 (i.e. a 5\% bias), the new diameter for \epsoph\ is
$\theta_\mathrm{UD}=2.895\pm0.005$\,mas, with a reduced $\chi^2$ of 2.5
instead of our result $\theta_\mathrm{UD}=2.888\pm0.003$\,mas, with a
reduced $\chi^2$ of 1.0. Not only our final result would be barely
affected, moreover the reduced $\chi^2$ becomes much larger, because
points calibrated by our large calibrator become completely inconsistent
with the rest of the batch.
\end{itemize}
Another possible source of bias, and possibly the ultimate one, is the
wavelength calibration. This has been done by \citet{merand-phd} and the
process leads to a wavelength calibration accuracy of 0.005\,$\mu$m. We
changed the software wavelength calibration by this amount and redid the
reduction and calibration. This led to a bias of 0.006\,mas. Hence, we
added quadratically a $0.006$\,mas error, leading to a final uncertainty
of $0.007$\,mas.

\begin{table}
\caption{
Squared visibility measurements obtained for \epsoph\ are given.  $B$ is
the projected baseline length, and ``PA" is the azimuth of the projected
baseline (counted positively from North to East).
\label{tab:visib-table}
}
\begin{tabular}{lrrc}
\hline \hline
MJD & $B$ (m) & PA\,($^\circ$) & $V^2 \pm \sigma(V^2)$ \\
\hline
\noalign{\smallskip}
53936.21641 &  154.375 & $-32.471$ & $0.03131 \pm 0.00173$\\
53936.25567 &  165.132 & $-36.878$ & $0.01450 \pm 0.00049$\\
53937.20561 &  151.978 & $-31.267$ & $0.03746 \pm 0.00104$\\
53937.23681 &  160.961 & $-35.346$ & $0.01901 \pm 0.00055$\\
53937.26478 &  167.893 & $-37.774$ & $0.01054 \pm 0.00028$\\
\hline
\end{tabular}
\end{table}

\subsection{Limb darkened angular diameter and photospheric radius 
\label{radius}}

In order to estimate the unbiased angular diameter from the measured
visibilities it is necessary to know the intensity distribution of the
light on the stellar disk, i.e., the limb darkening (LD). As we do not
fully resolve \epsoph, we cannot measure the limb darkening directly
from the data. We thus model it using the MARCS models
\citep{gustafsson08}\footnote{http://www.marcs.astro.uu.se/} for the
computation of the intensity profile of the star, taking into account
the actual spectral transmission function of the FLUOR instrument
\citep{merand-phd}.  The LD coefficients have been computed with the
TURBOSPECTRUM code \citep{alvarez98}.  The result is shown in
Fig.~\ref{limb-darkening}. It is to be noted that taking an intensity
profile from a different model, say the one predicted by the ATLAS9
model from Kurucz, using Claret's laws \citep{claret00}, results in the
same final result, within fractions of the statistical error bar.  The
reason is that for the relatively large spectral bandwidth of FLUOR and
considering that we measure first-lobe visibilities only, the difference
between the MARCS and ATLAS models is negligible (of the order of a
fraction of our diameter error bar).  The magnitude of the limb
darkening effect being much smaller in the infrared than in the visible,
our resulting limb darkened angular diameter measurement in the $K$ band
is largely unaffected by the choice of the limb darkening model.

\myfigure{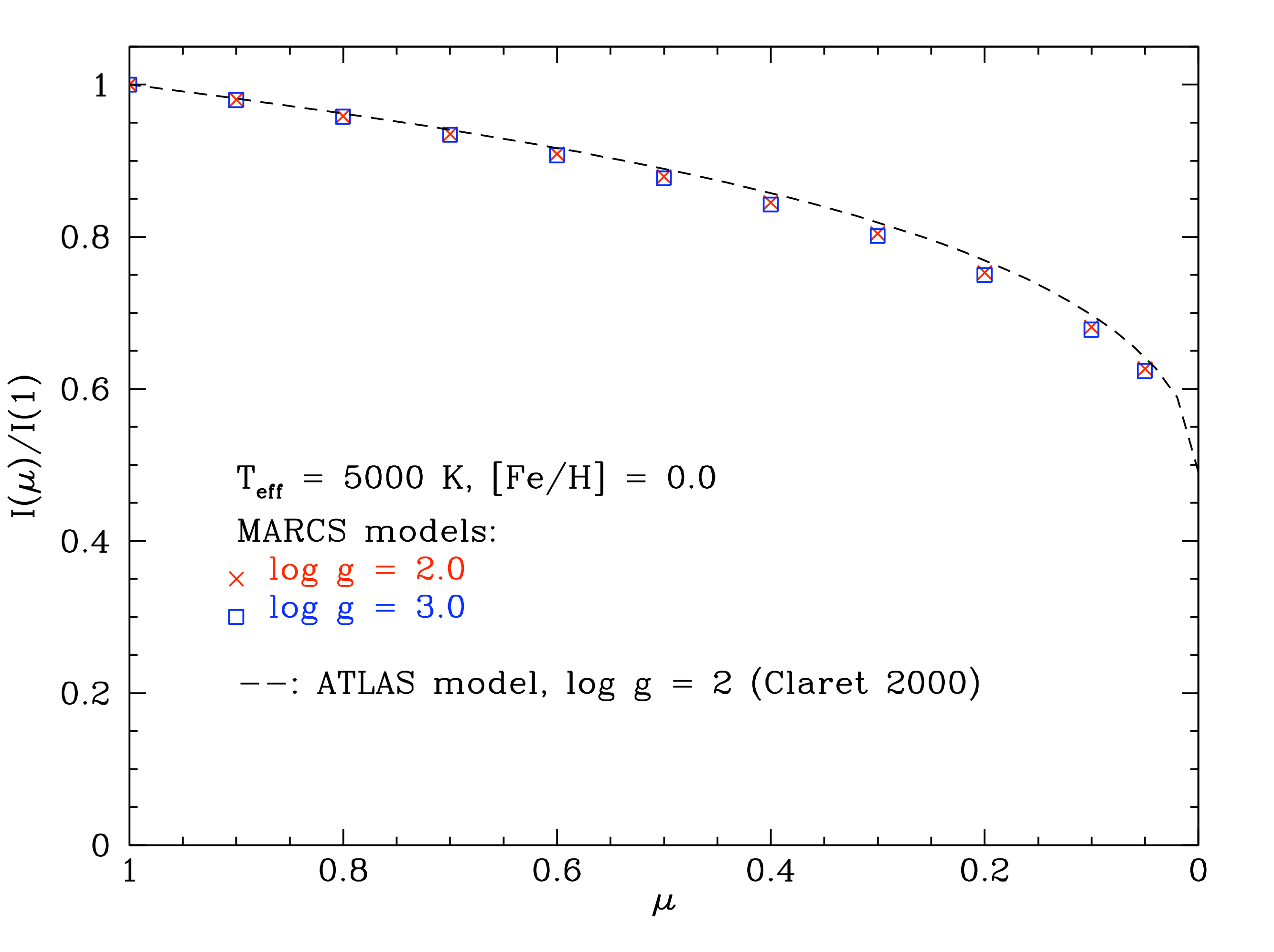}
{Comparison of the intensity profiles of \epsoph\ from the MARCS ({\it
red crosses} and {\it blue squares} for two values of $\log g$) and
Claret's~\citep{claret00} ({\it dashed curve}) model.}
{limb-darkening}
{}

The result of the visibility fit is presented in Fig.~\ref{visib-curve}
using the MARCS limb darkening model. We derive the following limb
darkened disk angular diameter:
\begin{equation}
\theta_{\rm LD} (\epsoph) = 2.961 \pm 0.003\ (\mathrm{stat.}) \pm 0.006\
(\mathrm{syst.})\ \mathrm{mas}
\end{equation}
\myfigure{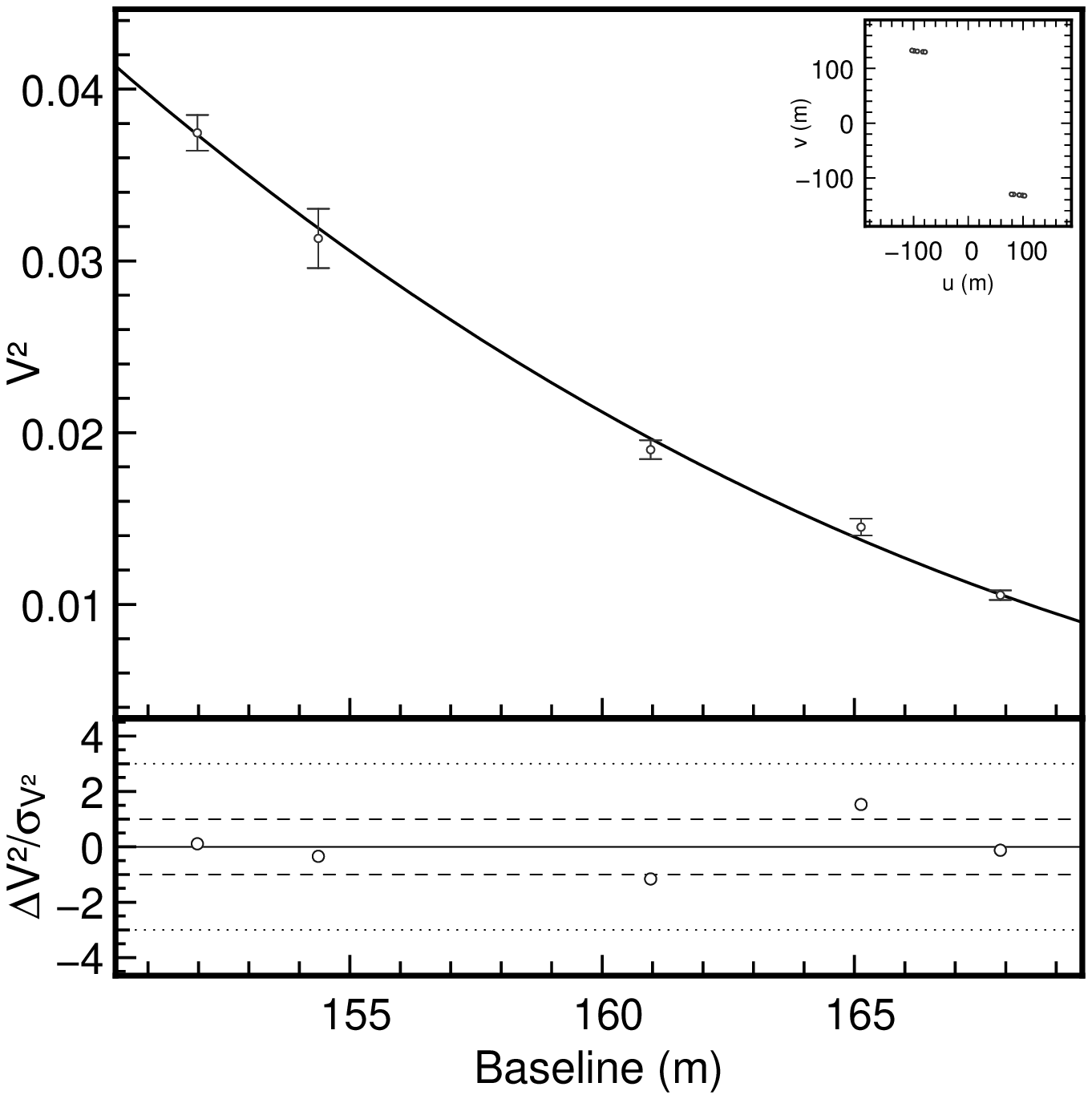}
{Squared visibilities and adjusted limb darkened disk visibility model
for \epsoph\ (see also Table~\ref{tab:visib-table}). The lower panel
shows the residuals to the model, with $1\sigma$ ({\it dashed line}) and
$3\sigma$ ({\it dotted line}) limits. The {\it inset} panel shows the
($u$, $v$) coverage. Note that the error bar given here is the formal
error bar: we added some systematics for final result (see text).}
{visib-curve}
{}
This value is compatible with the spectrophotometric angular diameter
estimate by \citet{cohen99} of $\theta_{\rm LD} = 3.00 \pm 0.03$\,mas.
We took \epsoph's parallax from the reprocessed \emph{Hipparcos}
catalogue by \citet{vanleeuwen07a, vanleeuwen07b}:
\begin{equation}
\pi(\epsoph) = 30.64 \pm 0.20\ {\rm mas}~(\pm 0.65\%).
\end{equation}
This value compares well with \citet{vanaltena95}, and the original
\emph{Hipparcos} catalogue \citep{esa97}, but is more precise.  We
finally derive the photospheric linear radius:
\begin{equation}
R(\epsoph) = 10.39 \pm 0.07 \rsun~(\pm 0.67\%).
\end{equation}
In spite of the the relatively high precision of the parallax, it is by
far the limiting factor for the precision of the radius.

\section{Discussion}
\label{sec:discuss}

In this study we have constructed stellar models of red giants in both
shell H-burning and core He-burning phases and compared their adiabatic
frequencies with the frequencies of \epsoph\ observed by the MOST
satellite and published by \citet{barban07}. We have also measured the
radius of the star through optical interferometry using the CHARA/FLUOR
instrument.

We have demonstrated that the observed frequencies of \epsoph\ are
consistent with radial $p$-mode pulsations of a red giant at the
relevant position on the \hrd. Unfortunately, the radial mode
frequencies cannot distinguish between the two phases of stellar
evolution -- shell H-burning and core He-burning. This is hardly
surprising, since the models inside the box on the \hrd\ in either phase
would have approximately similar radii, and the large separation depends
crucially on the radius of the star. However, the seismic information
helps us to constrain the radius of the star to a much narrower range
than that possible by the errorbox on the \hrd.

\citet{kal08a} have carried out a seismic modelling study of \epsoph\
also, but with important differences. Firstly, they have interpreted the
observed peaks in the power spectrum of \epsoph\ to be radial as well as
nonradial modes. They identify the  sharp narrow peaks in the power
spectrum as long-lived nonradial modes, as compared to the broad
Lorentzian envelopes, resulting in short-lived modes, which have been
identified as radial modes by \citet{barban07}. Secondly, they have
matched the observed frequency values, and not the large separations, to
the theoretical model frequencies. Lastly, they have used only shell
H-burning models. Given these differences, it is not surprising that
they obtained a slightly different result than ours.

Little is known whether $p$-modes, radial or  nonradial, can be excited
in red giants to an amplitude high enough to be observable. A detailed
theoretical study of models of $\alpha$~UMa, observed by
\citet{buzasi00} with the WIRE satellite, and of  similar 2\msun\ models
on the lower giant branch, by \citet{dziem01} provides some insight on
the oscillation characteristics of lower giant branch stars. Giant stars
are characterised by an inner cavity that can support gravity waves
($g$-modes), and an outer cavity that supports acoustic waves
($p$-modes).  Observable modes in the outer cavity are mixed modes, with
a $g$-mode character in the inner cavity and $p$-mode character in the
outer cavity. \citet{dziem01} showed that such mixed $p$-modes can have
substantial  amplitudes, and that low degree modes with $\ell=2,3$,
together with the radial $p$-modes can be unstable. According to their
models of lower giant branch stars, high amplitudes in the outer cavity
arise only for modes with $\ell=2$. The excitation properties of
$p$-modes in more luminous red giants lying around the middle of the red
giant branch, which have very deep convection zones, such as \epsoph,
are mostly unexplored.

A careful visual examination of the observed frequency spectrum of
\epsoph\ \citep{barban07} reveals a clear comb-like structure with
reasonably regular spacing of $\sim 5.3\,\muhz$. In an adiabatic
pulsation calculation of a theoretical model, a series of regularly
spaced radial modes accompanied by a dense forest of nonradial modes are
obtained (see Fig.~\ref{fig:nonradial}).  The observed frequencies of
\epsoph, as given by \citet{barban07}, can be matched reasonably well to
the radial modes. In principle, they can also be matched easily to many
of the closely spaced nonradial modes as well, but the question remains
as to why the majority of the nonradial modes are not observed. 

\myfigure{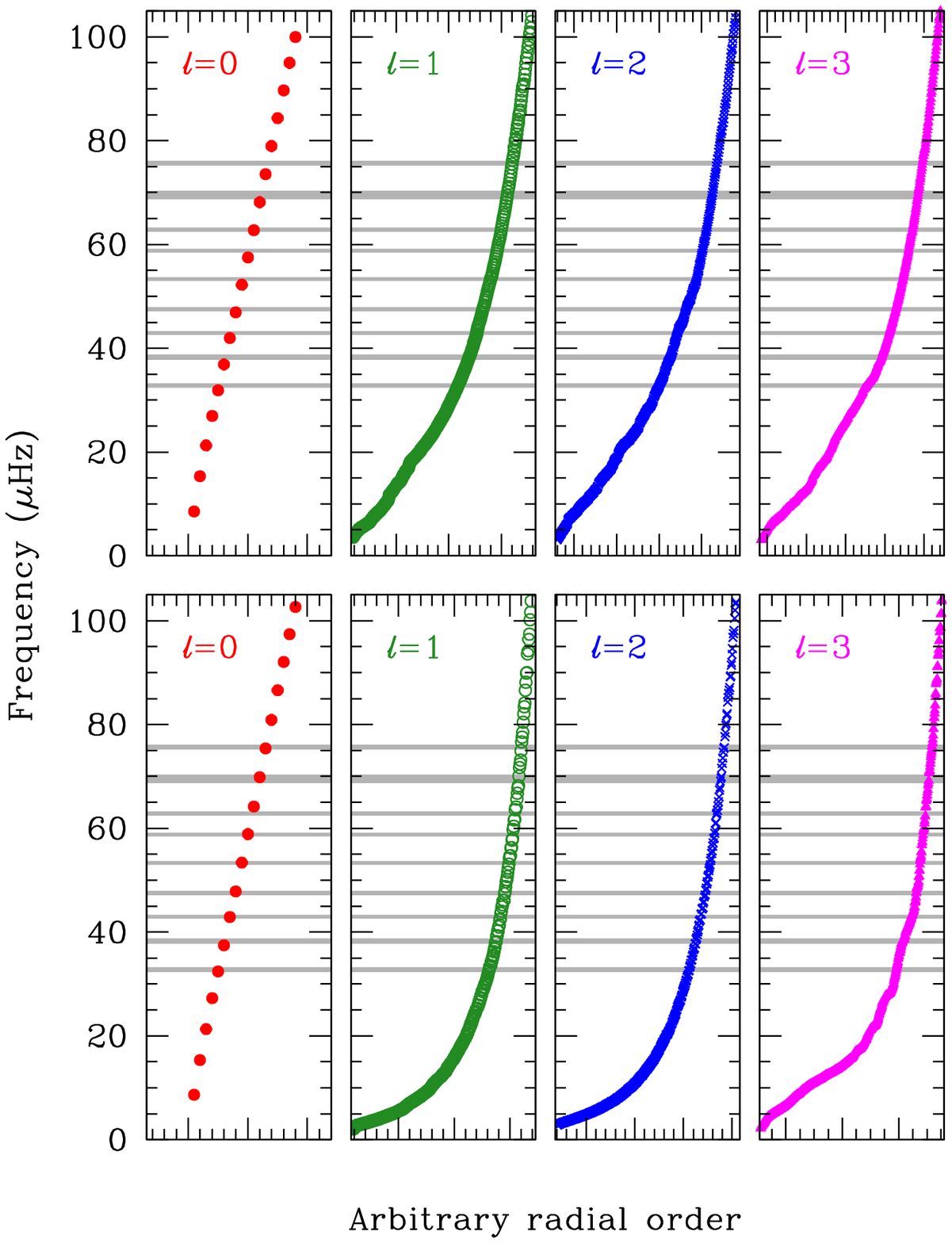}
{Adiabatic pulsation frequencies of radial ($\ell=0$) and nonradial
($\ell=1,2,3$) modes in a typical red giant model inside the \epsoph\
errorbox on the \hrd\ are shown. The frequencies of each degree are
plotted in increasing order of magnitude against some arbitrary
assignment of radial order (which can be simply treated as the serial
number of the mode in the eigenspectrum). The {\it top} row shows the
frequencies of a 1.8\msun\ star inside the \epsoph\ in the shell
H-burning phase. The {\it bottom} row shows the frequencies of the same
star when it again appears inside the \epsoph\ errorbox in the later
phase of stable core He-burning. The horizontal {\it grey} bands
represent the observed MOST frequencies of \epsoph\ with $\pm 1\sigma$
error bars.}
{fig:nonradial}
{}

A possible explanation of this may be provided in terms of the
normalised mode inertia $E$, defined according to \citet{jcd04} as
\begin{equation}
E = \frac{\int_V \rho |\vec{\xi} (r)|^2 \mathrm{d}V} 
{M |\vec{\xi}(R)|^2},
\end{equation}
where $\vec{\xi}$ is the displacement vector, and the integration is
carried out over the volume $V$ of the star.  Most nonradial modes are
strongly trapped in the core, with only a few modes being trapped in the
envelope. The modes trapped in the core have high inertia and therefore
unlikely to have enough surface amplitudes. On the other hand, the modes
trapped in the envelope can have inertia as low as the radial modes, and
therefore, be seen on the surface. Fig.~\ref{fig:inertia} shows the
normalised mode inertia for radial as well as nonradial modes in a
typical model inside the \epsoph\ errorbox. Only selected $\ell=1,2,3$
modes have inertia comparable to the radial modes. However, a reliable
estimate of the surface amplitude can only be obtained when nonadiabatic
effects in the outer layers are taken into account.
\myfigure{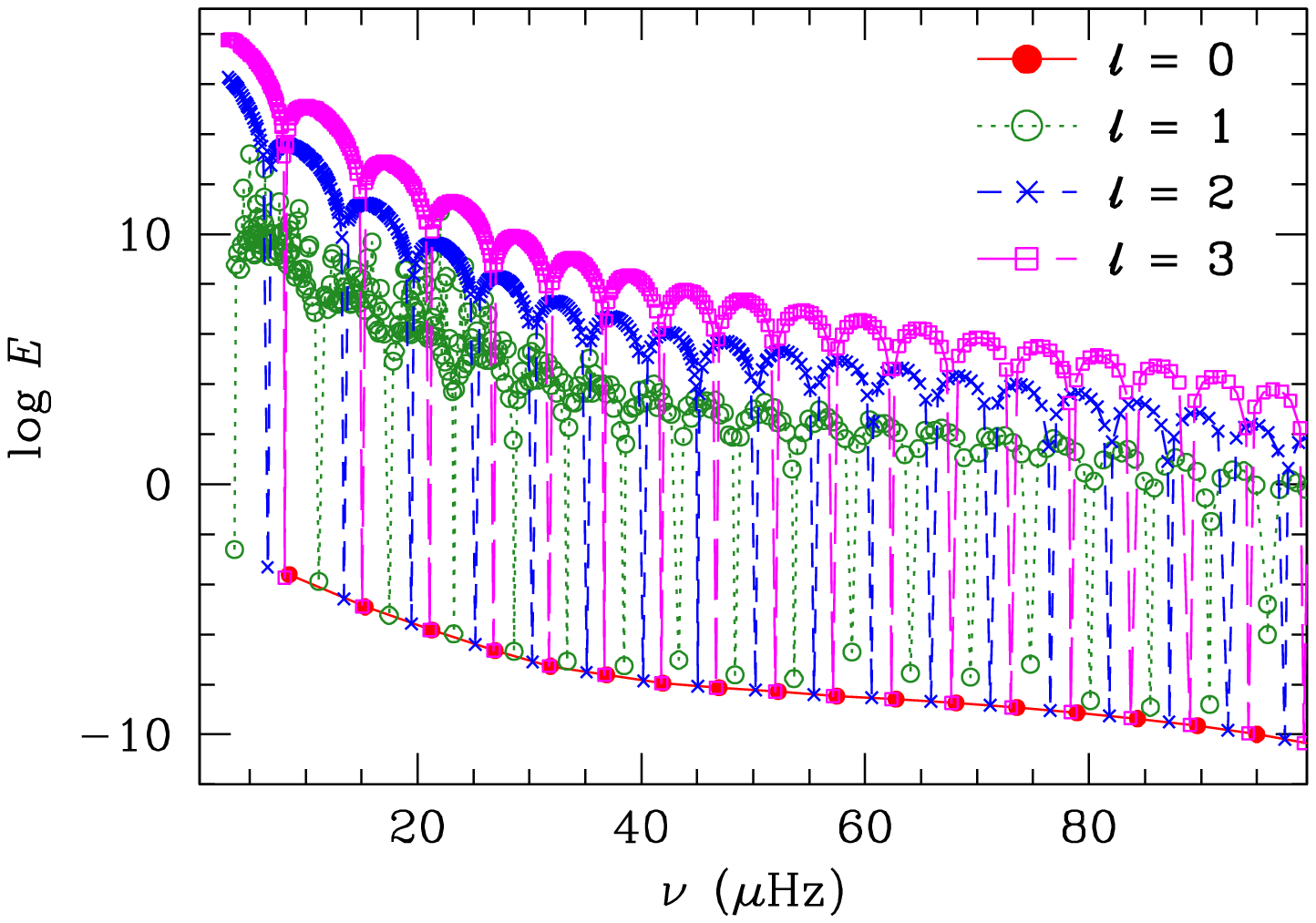}
{Normalised mode inertia of $\ell=0,1,2,3$ modes as a function of
frequency is shown for a $1.8\msun$ star in shell hydrogen burning phase
inside the \epsoph\ errorbox. The individual eigenfrequencies are joined
by lines for each degree ({\it red filled circles} with {\it solid}
lines for $\ell=0$; {\it green empty circles} with {\it dotted} lines
for $\ell=1$; {\it blue crosses} with {\it short dashed} lines for
$\ell=2$; {\it magenta empty squares} with {\it long dashed} lines for
$\ell=3$).  }
{fig:inertia}
{}

In a recent theoretical work involving nonadiabatic treatment of the
excitation mechanism, \citet{dupret09} have found that despite their low
amplitudes, a selection of nonradial modes may still have appreciable
heights in the power spectrum of red giants due to their long lifetimes,
and thus be detected in observations.  However, the detection of such
modes depends crucially on the evolutionary stage of the star on the red
giant branch.  Theoretical computations for intermediate red giant
branch stars like \epsoph\ predict much longer lifetimes for nonradial
modes ($\ga 50$ days) than radial modes ($\la 20$ days) because of their
larger inertia.  But if the duration of observation is shorter than
these lifetimes, as the case is for MOST observations of \epsoph, it is
not possible to resolve these nonradial modes in the power spectrum.
This implies that the modes attain smaller heights in the power spectrum
and they become extremely difficult to detect. 

Further, in the higher part of the frequency domain, the detectable
nonradial modes also appear with an asymptotic regular separation
pattern similar to that found in main sequence stars. This means that
the $\ell=0$ and $\ell=2$ modes will appear close to each other, with
the $\ell=1$ modes occurring roughly midway between them.  For \epsoph\
such a pattern implies that if we consider nonradial modes to be present
in the spectrum, the large separation (of $\ell=0$ modes, for example)
would be almost double the value than that obtained by postulating only
radial modes. Such a high value of the large separation ($\sim
11\,\muhz$) is inconsistent with the position of \epsoph\ on the \hrd.
However, according to \citet{dupret09}, the lifetimes of some $\ell=2$
modes which are strongly trapped in the envelope are comparable to that
of the radial ones.  So these envelope-trapped $\ell=2$ modes could
indeed be detectable. Trapping of $\ell=1$ modes is less efficient, as
found by \citet{dziem01} too, and hence they may not be observable.
However, in the absence of any detailed calculations for the mode
amplitudes and lifetimes for the specific case of \epsoph, in this work
we have adopted \citet{barban07}'s interpretation of the frequencies as
radial modes only. An alternative modelling analysis taking into account
the possibility of nonradial modes might lead to a different set of
model parameters, as found by \citet{kal08a}, for example. The
theoretical justification behind the presence of only a few specific
nonradial modes among the possible dense spectrum of such modes requires
further detailed study.

\begin{table*}
\caption{
Stellar parameters for the models with lowest \chisqdeltanu\ and
\chisqnu\ values in either shell H-burning or core He-burning phase that
have radii within $\pm 1\sigma$ of the interferometric radius are given.  
\label{tab:best-rad}
}
\begin{tabular}{clcccccccccc}
\hline
\hline
Minimised by & Evolutionary phase & $Y_0$ & $Z_0$ & $\alpha$ & $M/\msun$ & Age (Myr) 
& \lteff & $\log L/\lsun$ & $R/\rsun$ 
& \chisqnu  & \chisqdeltanu  \\
\hline
\chisqdeltanu\ or \chisqnu 
& Shell H-burning & 0.275 & 0.012 & 1.8 & 1.9 & \phantom{1}985 & 3.685 & 1.732 &
10.44  & 3.38 & 0.73\\
& Core He-burning & 0.270 & 0.013 & 1.8 & 1.8 & 1322 & 3.691 & 1.754 &
10.41  & 3.42 & 0.67\\
\hline
\end{tabular}
\end{table*}

Our interferometric measurements yield a value of the radius of \epsoph\
that is in close agreement with the radii of our best models obtained
through seismic analysis. The interferometric radius was not used as a
constraint to choose the best model, but was rather checked {\it a
posteriori} against the seismic values. The range of possible values of
radius of stellar models within the errorbox on the \hrd\ for \epsoph\
is $9\rsun$ to $12\rsun$. But the seismic information, specifically the
frequencies or the large separation, restricts the radius to a much
smaller range. The best-fitting models using large separation comparison
have radii of $R \approx 10.5 \rsun$ for both shell H-burning and core
He-burning phases. This value is within $2\sigma$ of the interferometric
radius. Even considering all models with $\chi^2 \leq 1$ constrains the
range of radius to $10.4\leq R/\rsun \leq 11.2$ for frequency
comparison, and to $10.2 \leq R/\rsun \leq 11.6$ for large separation
comparison (see Fig.~\ref{fig:chisq}). This range encompasses the much
narrower limit for the radius set by the interferometric measurements.
It is remarkable that despite the inherent uncertainties in the
modelling of the outer layers of a star, the seismic analysis alone
leads to a value of the radius that is in such good agreement with an
independent direct estimate of the radius.  The radius obtained by
\citet{kal08a} through frequency fitting, $R=10.8\rsun$, seems to be
more removed from the interferometric radius than our seismic value is,
although it is difficult to directly compare the two in view of the
absence of error bars for the former.

\myfigure{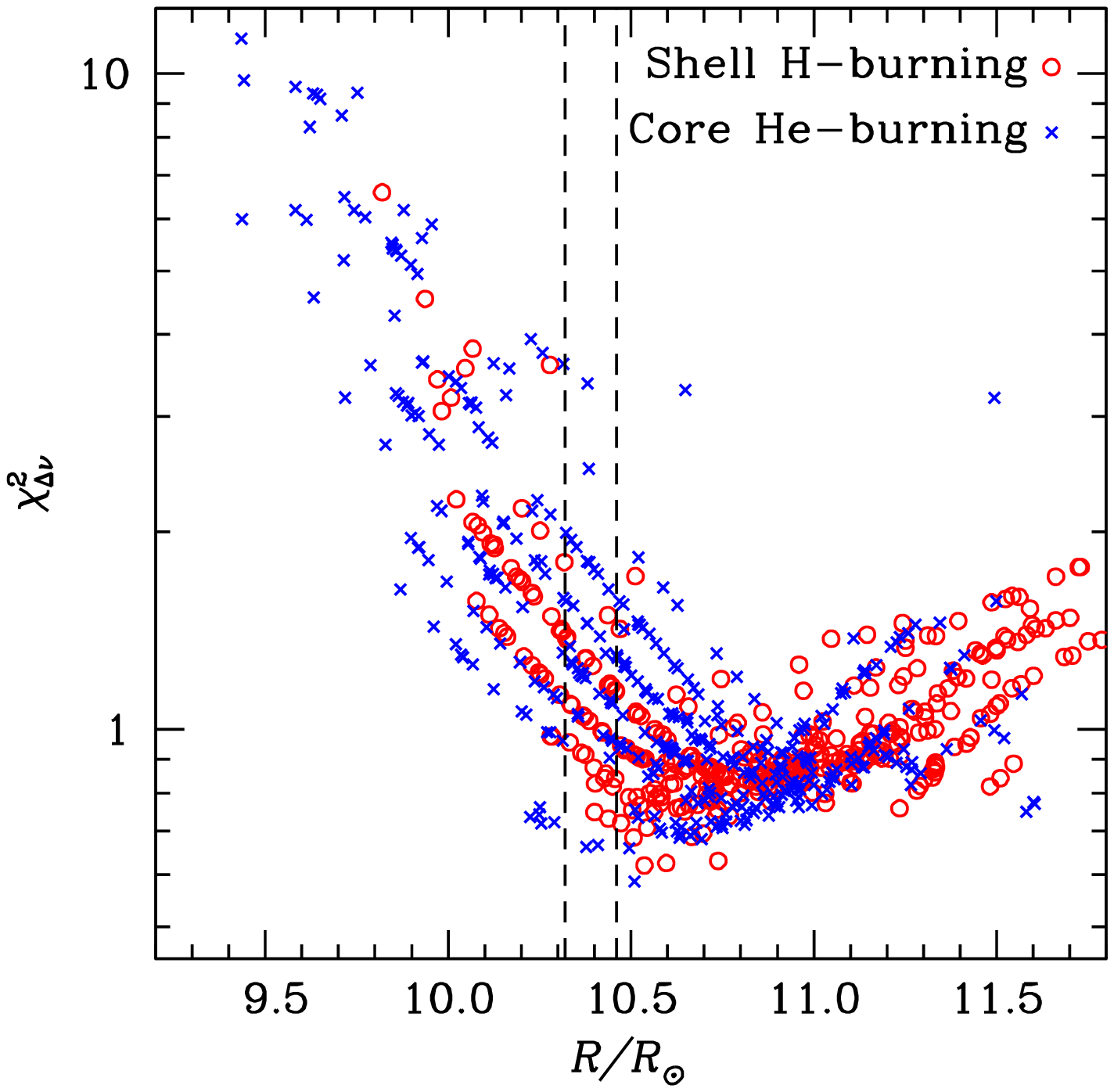}
{The \chisqdeltanu\ values are shown as functions of radius for all
stellar models constructed within the \epsoph\ errorbox.  The {\it red
circles} represent the shell H-burning and the {\it blue crosses}
represent the core He-burning models. The vertical {\it dashed} lines
denote the 1$\sigma$ interval for the interferometric radius.}
{fig:chisq}
{}

\myfigure{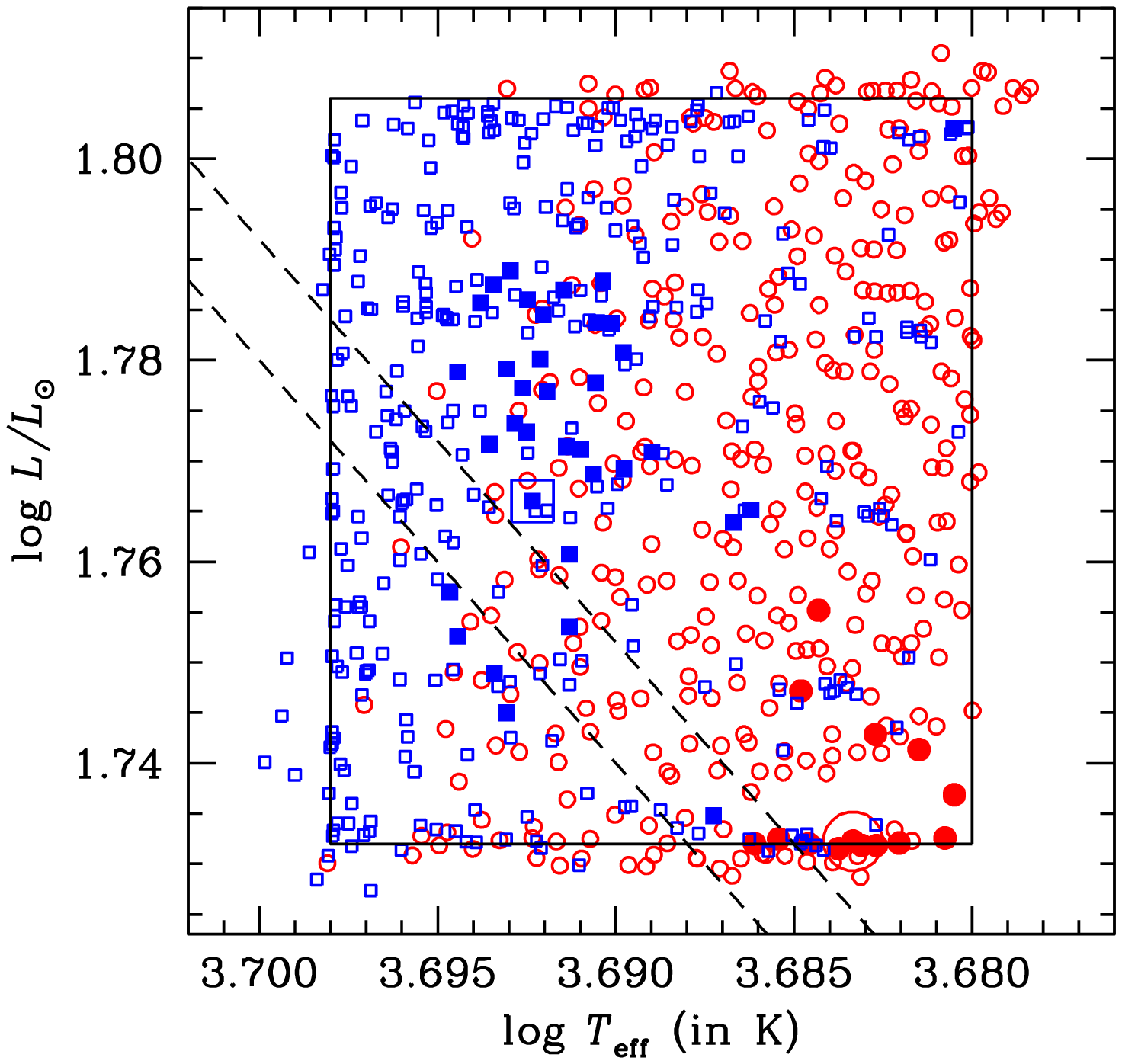}
{All the computed stellar models inside and around the \epsoph\ errorbox
on the \hrd\ are shown. The {\it red open circles} represent the shell
H-burning and the {\it blue open squares} represent the core He-burning
models. The {\it filled} symbols represent the models with
$\chisqdeltanu \leq 0.8$ for each phase of evolution. The {\it big
circle} and the {\it big square} show the best models in the two phases.
The {\it solid} line rectangle represents the adopted errorbox of
\epsoph. The {\it dashed} lines denote the loci of constant radii
$R=10.32\rsun$ and $R=10.46\rsun$, which are the $\pm 1\sigma$ bounds
for the interferometric radius of \epsoph.}
{fig:hrd-rad}
{}

As mentioned above, in the present study the independent estimate of the
radius was not used as an additional constraint for choosing the best
seismic model. However, since the modelling is in no way influenced by
the presence of the radius information, we can check what would be the
result if indeed the radius is used as a constraint. We use the
$1\sigma$ interval of the radius to restrict the position of the star on
the \hrd, along with the adopted values of luminosity and effective
temperature. This leads to a trapezoidal area on the \hrd\ (see
Fig.~\ref{fig:hrd-rad}) as the errorbox for \epsoph. The parameters of
the models inside this smaller errorbox that fit the seismic data best
are shown in Table~\ref{tab:best-rad}. In this case, the minimisations
according to frequencies and large separations yield the same best
models in either phase of evolution. The minimum values of
\chisqdeltanu\ are marginally higher than in the more general case (cf.\
Table~\ref{tab:bestfit}).  However, the minimum \chisqnu\ values are
significantly higher, indicating that although the observed large
separations are quite well matched by these models, the absolute
frequency values are somewhat shifted. This is wholly expected since the
large separations are strongly influenced by the radius, even if the
frequencies themselves may be shifted due to inadequate modelling of the
surface effects. In other words, restricting the radius value implies a
strong constraint on the large separation, but not necessarily on the
absolute frequencies.  This is also borne out in the more general case
(when the radius constraint is not used) in Fig.~\ref{fig:hrd-rad} where
the models with lowest \chisqdeltanu\ values ($\leq 0.8$) in either
phase of evolution lie in a broad band almost parallel to the
interferometric radius band, indicating a constant higher radius value
($\sim 10.55 \pm 0.15 \rsun$) common to all of them. The seismic values
of the mass and radius of \epsoph\ yield an average large separation
value of $5.35\pm 0.18\,\muhz$ according to the scaling formula of
\citet{kb95}, which is completely consistent with the current data
\citep{barban07}.

It is also possible to determine the effective temperature of the star
from its measured radius and photometric data. From
Fig.~\ref{fig:hrd-rad} itself, it is evident that the intersection of
the luminosity limits and the interferometric radius ranges indicates a
temperature range of $\lteff = 3.695 \pm 0.010$ ($4955 \pm 100$\,K)
which is consistent with our adopted range of effective temperature and
has a similar uncertainty. Further, a fit of the available photometric
data on \epsoph\ ($BVJHK$ bands) using tabulated Kurucz models yields
$\lteff = 3.691 \pm 0.002$ ($4912 \pm 25$\,K) for surface gravity values
typical of stars in the relevant zone of the \hrd. Actually, a change of
$0.1$ dex in $\log g$ makes a difference of only 1\,K in \teff, while
the uncertainty in the measured angular diameter contributes about 4\,K
in the error estimate. Again, these value of \teff\ are completely
contained in the range that we have used from \citet{deridder06}.  Thus
our adopted values of $L$ and \teff\ are consistent with the independent
measurements of parallax ({\it Hipparcos}) and the angular diameter
(this paper). 

It is evident from this study that the seismic information alone can go
a long way in constraining the most important stellar parameters of red
giants. Even with a very limited data set, it was possible to obtain a
reasonably narrow range of parameters for \epsoph, and the radius
estimate from the seismic modelling stands in close agreement with a
completely independent interferometric measurement. However, the
accuracy of the models can be greatly enhanced by the additional
information about the interferometric radius. An independent radius
measurement, with a high precision such as provided by interferometry,
helps in reducing the size of the errorbox on the \hrd, making the task
of searching for the best model easier. This is the first instance of
the coming together of asteroseismology and interferometry for red giant
stars, and clearly illustrates the huge potential of this combination in
detailed studies of such stars.

\begin{acknowledgements}

The authors would like to thank all the CHARA Array and Mount Wilson
Observatory day-time and night-time staff for their support.  The CHARA
Array was constructed with funding from Georgia State University, the
National Science Foundation, the W.~M.\ Keck Foundation, and the David
and Lucile Packard Foundation. The CHARA Array is operated by Georgia
State University with support from the College of Arts and Sciences,
from the Research Program Enhancement Fund administered by the Vice
President for Research, and from the National Science Foundation under
NSF Grant AST~0606958.  This work also received the support of PHASE,
the high angular resolution partnership between ONERA, Observatoire de
Paris, CNRS and University Denis Diderot Paris 7.  This research took
advantage of the SIMBAD and VIZIER databases at the CDS, Strasbourg
(France), and NASA's Astrophysics Data System Bibliographic Services.
Part of this work was supported by the Research Fund of K.~U.\ Leuven
under grant GOA/2003/04 for AM and CB. STR acknowledges partial support
by NASA grant NNH09AK731. The authors thank Marc-Antoine Dupret and
Sarbani Basu for their valuable comments and suggestions.

\end{acknowledgements}


\begin{thebibliography}{}

\bibitem[Alvarez \& Plez(1998)]{alvarez98} 
Alvarez, R., \& Plez, B.\ 
1998, \aap, 330, 1109

\bibitem[Bahcall \& Pinsonneault(1992)]{bahcall92}
Bahcall, J.~N.\ \& Pinsonneault, M.~H.\ 
1992, Rev.\ Mod.\ Phys., 64, 885

\bibitem[Barban \ea(2007)]{barban07} 
Barban, C., \ea\
2007, \aap, 468, 1033 

\bibitem[B{\"o}hm-Vitense(1958)]{bohm58} 
B{\"o}hm-Vitense, E.\ 
1958, \zap, 46, 108 

\bibitem[Buzasi \ea(2000)]{buzasi00}
Buzasi, D., Catanzarite,  J., Laher, R., Conrow, T., Shupe, D., 
Gatier, T.~N.~III, Kreidl, T.\ \& Everett, D.\ 
2000, \apj, 532, L133

\bibitem[Cayrel de Strobel \ea(2001)]{cay01}
Cayrel de Strobel, G., Soubiran, C., \& Ralite ,N.\ 
2001, \aap, 373, 159

\bibitem[Christensen-Dalsgaard(2004)]{jcd04}
Christensen-Dalsgaard, J.\ 
2004, \solphys, 220, 137

\bibitem[Claret(2000)]{claret00} 
Claret, A.\ 
2000, \aap, 363, 1081

\bibitem[Clayton(1968)]{clayton68} 
Clayton, D.~D.\ 
1968, Principles of Stellar Evolution and Nucleosynthesis, 
New York: McGraw-Hill (1968)

\bibitem[Cohen \ea(1999)]{cohen99} 
Cohen, M., Walker, R.~G., Carter, B., \ea\ 
1999, \aj, 117, 1864

\bibitem[Cole \& Deupree(1983)]{col83}
Cole, P.~W. \& Deupree, R.~G.\ 
1983, \apj, 269, 676 

\bibitem[Coud\'e du Foresto \ea(1997)]{coude97} 
Coud\'e du Foresto, V., Ridgway, S., Mariotti, J.-M.\  
1997, \aaps, 121, 379

\bibitem[Coud\'e du Foresto \ea(2003)]{coude03} 
Coud\'e du Foresto, V., Bord\'e, P., M\'erand, A., \ea\ 
2003, Proc. SPIE, 4838, 280

\bibitem[Creevey \ea(2007)]{creevey07} 
Creevey, O.~L., Monteiro, M.~J.~P.~F.~G., Metcalfe, T.~S., \ea\ 
2007, \aap, 659, 616

\bibitem[Cunha \ea(2007)]{cunha07} 
Cunha M.~S., Aerts C., Christensen-Dalsgaard J., Baglin A., \ea\ 
2007, \aapr, 14, 217

\bibitem[de Ridder \ea(2006)]{deridder06}
de Ridder, J., Barban, C., Carrier, F., Mazumdar, A., Eggenberger, P., 
Aerts, C., Deruyter, S., \& Vanautgaerden, J.\ 
2006, \aap, 448, 689 

\bibitem[Demarque \& Mengel(1971)] {dem71}
Demarque, P. \& Mengel, J.~G.\ 
1971, \apj, 164, 317

\bibitem[Demarque \ea (2004)] {dem04}
Demarque, P. , Woo, J.-H., Kim, Y.-C. \& Yi, S.~K.\ 
2004, \apjs, 155, 667

\bibitem[Demarque \ea(2008)] {yrec}
Demarque, P., Guenther, D.~B., Li, L.~H., Mazumdar, A., \& Straka, C.~W.\
2008, \apss, 316, 31  

\bibitem[Dupret \ea(2009)]{dupret09}
Dupret, M.-A., Belkacem, K., Samadi, R.\ \ea\ 
2009, \aap, accepted (doi: 10.1051/0004-6361/200911713)

\bibitem[Dziembowski \ea(2001)]{dziem01} 
Dziembowski, W.~A., Gough, D.~O., Houdek, G., \& Sienkiewicz, R.\ 
2001, \mnras, 328, 601 

\bibitem[ESA(1997)]{esa97} 
ESA
1997, The Hipparcos and Tycho Catalogues, ESA SP-1200

\bibitem[Ferguson \ea(2005)]{ferg05}
Ferguson, J.~W., Alexander, D.~R., Allard, F., Barman, T., 
Bodnarik, J.~G., Hauschildt, P.~H., Heffner-Wong, A., \& Tamanai, A.\ 
2005, \apj, 623, 585

\bibitem[Frandsen \ea(2002)]{frandsen02} 
Frandsen, S., Carrier, F., Aerts, C.\ \ea\ 
2002, \aap, 394, L5 

\bibitem[Grevesse \& Sauval(1998)]{gs98} 
Grevesse, N., \& Sauval, A.~J.\ 
1998, \ssr, 85, 161

\bibitem[Guenther(1994)]{jig}
Guenther, D.~B.\
1994, \apj, 422, 400 

\bibitem[Guenther \ea(1992)]{gue92}
Guenther, D.~B., Demarque, P.\, Kim, Y.-C., \& 
Pinsonneault, M.~H.\
1992, \apj, 87, 372

\bibitem[Gustafsson \ea(2008)]{gustafsson08} 
Gustafsson, B., Edvardsson B., Eriksson K., \ea\ 
2008, A\&A, 486, 951

\bibitem[Hekker \ea(2008)]{hekker08} 
Hekker, S., Barban, C., Kallinger, T., Weiss, W., de Ridder, J., 
Hatzes, A., \& COROT Team 
2008, Communications in Asteroseismology, 157, 319 

\bibitem[Iglesias \& Rogers(1996)]{opalop95}
Iglesias C.~A.\ \& Rogers F.~J.\ 
1996, \apj, 464, 943

\bibitem[Itoh \ea(1989)]{itoh89}
Itoh, N., Adachi, T., Nakagawa, M., Kohyama, Y., Munakata, H.\ 
1989, \apj, 339, 354

\bibitem[Kallinger \ea(2008a)]{kal08a}
Kallinger, T., Guenther, D.~B., Matthews, J.~M. \ea\
2008a, \aap, 478, 497

\bibitem[Kallinger \ea(2008b)]{kal08b} 
Kallinger, T., \ea\ 
2008b, Communications in Asteroseismology, 153, 84

\bibitem[Kervella \ea(2004)Kervella, S\'egransan,\& Coud\'e du Foresto]{kervella04} 
Kervella, P., S\'egransan, D. \& Coud\'e du Foresto, V.\ 
2004, A\&A, 425, 1161

\bibitem[Kervella \ea(2008)]{kervella08} 
Kervella, P., M\'erand, A., Pichon, B., \ea\ 
2008, A\&A, 488, 667

\bibitem[Kjeldsen \& Bedding(1995)]{kb95}
Kjeldsen, H., \& Bedding, T.~R.\ 
1995, \aap, 293, 87

\bibitem[Lee \& Demarque (1990)]{lee90}
Lee, Y.-W. \& Demarque, P.\ 
1990, \apjs, 73, 709

\bibitem[M\'erand(2005)]{merand-phd}  
M\'erand, A., Ph.D. thesis
2005, Universit\'e Paris 7.

\bibitem[M\'erand \ea(2005)M\'erand, Bord\'e \& Coud\'e du Foresto]{merand05} 
M\'erand, A., Bord\'e, P., \& Coud\'e du Foresto, V.\ 
2005, \aap, 433, 1155

\bibitem[M\'erand(2006)]{merand06} 
M\'erand, A., Coud\'e du Foresto, V., Kellerer, A., \ea\
2006, Proc. SPIE, 6268, 46

\bibitem[Moc\`{a}k \ea(2008)] {moc08}
Moc\`{a}k, M. ,M\"{u}ller, E., Weiss, A., \& Kiforidis, K.\ 
2008, \aap, 490, 265

\bibitem[Perrin(2003)]{perrin03} 
Perrin, G.\ 
2003, \aap, 400, 1173

\bibitem[Piersanti \ea(2004)]{ptc04}
Piersanti, L., Tornamb{\'e}, A., \& Castellani, V.\
2004, \mnras, 353, 243 

\bibitem[Reimers(1977)]{reimers77}
Reimers, D.\ 
1977, \aap, 57, 395

\bibitem[Rogers \& Nayfonov(2002)]{opaleos05}
Rogers, F.~J.\ \& Nayfonov, A.\ 
2002, \apj, 576, 1064

\bibitem[Schwarzschild \& H\"{a}rm (1962)] {sch62}
Schwarzschild, M. \& H\"{a}rm, R.\ 
1962, \apj, 136, 158 

\bibitem[Straka \ea(2007)]{straka07} 
Straka, C.~W., Demarque, P., \& Robinson, F.~J.\ 
2007, IAU Symposium, 239, 388

\bibitem[Sweigart (1987)] {swe87}
Sweigart, A.~V.\ 
\apjs, 65, 95

\bibitem[Sweigart \ea(1989)] {swe89}
Sweigart, A.~V., Greggio, L. \& Renzini, A.\
1989,\apjs, 69, 911  

\bibitem[ten Brummelaar(2005)]{tenbrummelaar05} 
ten Brummelaar, T.~A., McAlister, H.~A., Ridgway, S.~T., \ea\ 
2005, \apj, 628, 453

\bibitem[van Altena \ea(1995)van Altena, Lee \& Hoffleit]{vanaltena95} 
van Altena, W.~F., Lee, J.~T., \& Hoffleit, E.~D.\ 
1995, {The General Catalogue of Trigonometric Stellar Parallaxes}, 
4$^{\rm th}$ Edition, Yale University Observatory

\bibitem[van Leeuwen(2007a)]{vanleeuwen07a} 
van Leeuwen, F.\ 
2007a, {Hipparcos, the New Reduction of the Raw Data}, 
Astrophysics and Space Science Library, vol. 350, Springer

\bibitem[van Leeuwen(2007b)]{vanleeuwen07b} 
van Leeuwen, F.\ 
2007b, \aap, 474, 653

\bibitem[Walker \ea(2003)]{walker03} 
Walker, G., Matthews, J., Kuschnig, R.\ \ea\ 
2003, \pasp, 115, 1023

\bibitem[Yi \ea (2003)] {yi03}
Yi, S.~K., Kim, Y.-C., \& Demarque, P.\ 
2003, \apjs,  144, 259


\end{thebibliography}
\end{document}